\documentclass[author-year, prd, amsmath, amssymb, longbibliography, reprint, superscriptaddress, groupaddress, groupeaddress]{revtex4-1}

\usepackage{graphicx}
\usepackage{subfig}
\usepackage[export]{adjustbox}
\usepackage[table]{xcolor}
\usepackage{amssymb}
\usepackage{amsmath}
\usepackage{hyperref}
\usepackage{lipsum}
\usepackage{soul}
\usepackage{bbm}

\def\aj{\ref@jnl{AJ}}                   
\def\actaa{\ref@jnl{Acta Astron.}}      
\def\araa{\ref@jnl{ARA\&A}}             
\def\apj{\ref@jnl{ApJ}}                 
\def\apjl{\ref@jnl{ApJ}}                
\def\apjs{\ref@jnl{ApJS}}               
\def\ao{\ref@jnl{Appl.~Opt.}}           
\def\apss{\ref@jnl{Ap\&SS}}             
\def\aap{\ref@jnl{A\&A}}                
\def\aapr{\ref@jnl{A\&A~Rev.}}          
\def\aaps{\ref@jnl{A\&AS}}              
\def\azh{\ref@jnl{AZh}}                 
\def\baas{\ref@jnl{BAAS}}               
\def\bac{\ref@jnl{Bull. astr. Inst. Czechosl.}}  
\def\caa{\ref@jnl{Chinese Astron. Astrophys.}} 
\def\cjaa{\ref@jnl{Chinese J. Astron. Astrophys.}} 
\def\icarus{\ref@jnl{Icarus}}           
\def\jcap{\ref@jnl{J. Cosmology Astropart. Phys.}} 
\def\jrasc{\ref@jnl{JRASC}}             
\def\memras{\ref@jnl{MmRAS}}            
\def\mnras{\ref@jnl{MNRAS}}             
\def\na{\ref@jnl{New A}}                
\def\nar{\ref@jnl{New A Rev.}}          
\def\pra{\ref@jnl{Phys.~Rev.~A}}        
\def\prb{\ref@jnl{Phys.~Rev.~B}}        
\def\prc{\ref@jnl{Phys.~Rev.~C}}        
\def\prd{\ref@jnl{Phys.~Rev.~D}}        
\def\pre{\ref@jnl{Phys.~Rev.~E}}        
\def\prl{\ref@jnl{Phys.~Rev.~Lett.}}    
\def\pasa{\ref@jnl{PASA}}               
\def\pasp{\ref@jnl{PASP}}               
\def\pasj{\ref@jnl{PASJ}}               
\def\rmxaa{\ref@jnl{Rev. Mexicana Astron. Astrofis.}} 
\def\qjras{\ref@jnl{QJRAS}}             
\def\skytel{\ref@jnl{S\&T}}             
\def\solphys{\ref@jnl{Sol.~Phys.}}      
\def\sovast{\ref@jnl{Soviet~Ast.}}      
\def\ssr{\ref@jnl{Space~Sci.~Rev.}}     
\def\zap{\ref@jnl{ZAp}}                 
\def\nat{\ref@jnl{Nature}}              
\def\iaucirc{\ref@jnl{IAU~Circ.}}       
\def\aplett{\ref@jnl{Astrophys.~Lett.}} 
\def\apspr{\ref@jnl{Astrophys.~Space~Phys.~Res.}} 
\def\bain{\ref@jnl{Bull.~Astron.~Inst.~Netherlands}}  
\def\fcp{\ref@jnl{Fund.~Cosmic~Phys.}}  
\def\gca{\ref@jnl{Geochim.~Cosmochim.~Acta}}   
\def\grl{\ref@jnl{Geophys.~Res.~Lett.}} 
\def\jcp{\ref@jnl{J.~Chem.~Phys.}}      
\def\jgr{\ref@jnl{J.~Geophys.~Res.}}    
\def\jqsrt{\ref@jnl{J.~Quant.~Spec.~Radiat.~Transf.}} 
\def\memsai{\ref@jnl{Mem.~Soc.~Astron.~Italiana}} 
\def\nphysa{\ref@jnl{Nucl.~Phys.~A}}   
\def\physrep{\ref@jnl{Phys.~Rep.}}   
\def\physscr{\ref@jnl{Phys.~Scr}}   
\def\planss{\ref@jnl{Planet.~Space~Sci.}}   
\def\procspie{\ref@jnl{Proc.~SPIE}}   

\makeatletter
\renewcommand\@makecaption[2]{%
  \par
  \vskip\abovecaptionskip
  \begingroup
   \small\rmfamily
    \begingroup
     \samepage
     \flushing
     \let\footnote\@footnotemark@gobble
     \@make@capt@title{#1}{#2}\par
    \endgroup
  \endgroup
  \vskip\belowcaptionskip
}
\makeatother

\begin{document}

\title{Remarks on the maximum luminosity} 

\author{Vitor Cardoso}
\affiliation{Centro de Astrof{\'{i}}sica e Gravita{\c{c}}{\~{a}}o -- CENTRA, Departamento de F\'{\i}sica, Instituto Superior T{\'{e}}cnico -- IST, Universidade de Lisboa -- UL, Av.\ Rovisco Pais 1, 1049-001 Lisboa, Portugal}
\affiliation{Perimeter Institute for Theoretical Physics, 31 Caroline Street, North Waterloo, Ontario N2L 2Y5, Canada}

\author{Taishi Ikeda} 
\affiliation{Department of Physics, Graduate School of Science, Nagoya University, Nagoya 464-6602, Japan}

\author{Christopher J. Moore}
\email{christopher.moore@tecnico.ulisboa.pt}
\affiliation{Centro de Astrof{\'{i}}sica e Gravita{\c{c}}{\~{a}}o -- CENTRA, Departamento de F\'{\i}sica, Instituto Superior T{\'{e}}cnico -- IST, Universidade de Lisboa -- UL, Av.\ Rovisco Pais 1, 1049-001 Lisboa, Portugal}
\affiliation{DAMTP, Centre for Mathematical Sciences, University of Cambridge, Wilberforce Road, Cambridge, CB3 0WA, UK}

\author{Chul-Moon Yoo}
\affiliation{Department of Physics, Graduate School of Science, Nagoya University, Nagoya 464-6602, Japan}

\date{\today}

\begin{abstract}
The quest for fundamental limitations on physical processes is old and venerable. 
Here, we investigate the maximum possible power, or luminosity, that any event can produce.
We show, via full nonlinear simulations of Einstein's equations, that there exist initial conditions which give rise to arbitrarily large luminosities. 
However, the requirement that there is no past horizon in the spacetime seems to limit the luminosity to below the Planck value, ${{\cal L}_\textrm{P}\!=\!c^5/G}$.
Numerical relativity simulations of critical collapse yield the largest luminosities observed to date, ${\approx \! 0.2 {\cal L}_\textrm{P}}$. 
We also present an analytic solution to the Einstein equations which seems to give an unboundedly large luminosity; this will guide future numerical efforts to investigate super-Planckian luminosities.
\end{abstract}

\maketitle


%
\section{Introduction}
The boundaries of our Universe, and of the theories used to describe it, exert a strange fascination upon humankind. 
We have long wondered how old or large the Universe is. 
These questions led eventually to precise cosmological models, an active area of current research.
In the early 20th century, thought experiments regarding the fundamental limiting velocity laid the foundations of the special theory of relativity.
Shortly afterwards, quantum mechanics both answered and was made more complete by an understanding of the uncertainty principle, which provides intrinsic limitations on the measurements of ``position'' and ``velocity''. 

Gravitation, and specifically black holes (BHs) seem to play a central role in some of these issues:
point particles in General Relativity do not exist, and the compactness of systems is bounded by the BH limit~\cite{Thorne:1972,Choptuik:2009ww}. 
BHs also represent a limit on short-distance, high-energy physics~\cite{tHooft:1987vrq,Amati:1987wq,Banks:1999gd}. 
Surprisingly, BH physics is also intrinsically related to how ideal a fluid can be~\cite{Kovtun:2004de}, and even limits the accuracy of clocks~\cite{AmelinoCamelia:1994vs,Ng:1994zk,Gambini:2004de}.

\section{The maximum luminosity}
\begin{table}[t]
\begin{tabular}{|c|ll|}
\hline
\hline
Event  &  \multicolumn{2}{c|}{Peak luminosity}  \\
\hline
Three Gorges dam \cite{ThreeGorgesDam}           & $\;\;\sim\!10^{10}\,\textrm{W}$                            &$\;\;$($3\times 10^{-43}$)$\;\;$  \\
Most powerful laser \cite{MostPowerfulLaser}     & $\;\;\sim\!10^{15}\,\textrm{W}$                           &$\;\;$($3\times 10^{-38}$)$\;\;$  \\
Tsar Bomba \cite{TsarBomba}                      & $\;\;\sim\!10^{26}\,\textrm{W}$                           &$\;\;$($3\times 10^{-27}$)$\;\;$  \\
Solar luminosity \cite{SolarLuminosity}          & $\;\;\sim\!3\times 10^{26}\,\textrm{W}$                   &$\;\;$($10^{-30}$)$\;\;$  \\
$\gamma$-ray bursts \cite{LuminousGRB}           & $\;\;\sim\!3\times 10^{47}\,\textrm{W}$                   &$\;\;$($10^{-5}$)$\;\;$  \\
Inspiralling BHs \cite{Keitel:2016krm}           & $\;\;\sim\!7\times 10^{49}\,\textrm{W}$                   &$\;\;$($2\times 10^{-3}$)$\;\;$  \\
$\;\;$High-energy BH collision \cite{Sperhake:2009jz}$\;\;$  & $\;\;\sim\!7\times 10^{50}\,\textrm{W}$                   &$\;\;$($2\times 10^{-2}$)$\;\;$  \\
\rowcolor{lightgray} Critical collapse           & $\;\;\sim\!7\times 10^{51}\,\textrm{W}$                   &$\;\;$($2\times 10^{-1}$)$\;\;$ \\
End point of BH evaporation                      & $\;\;\sim\!3\times 10^{52}\,\textrm{W}$                    & $\;\;$($10^{0}$)$\;\;$ \\
\hline
\hline
\end{tabular}    
\caption{\label{tab:luminosity}
Approximate peak luminosities for some known events, in parenthesis we list the corresponding number in units of ${\cal L}_\textrm{P}$. 
Even if the solar luminosity is multiplied by the number of stars in the Universe (${\approx \! 10^{23}}$ \cite{HowManyStars}) the resultant luminosity is still only ${\approx \! 10^{-4}{\cal L}_\textrm{P}}$.
The inspiralling BH luminosity is for an equal mass binary on a quasi-circular orbit, whereas the high-energy BH collision is for BHs with initial velocities $0.75c$ and with a fine-tuned impact parameter.
The highlighted row is the luminosity associated with the spherically symmetric critical collapse of a scalar field as considered by \cite{PhysRevLett.70.9}; the luminosity of this process is calculated here for the first time.
The end point of BH evaporation refers to the final stage of the Hawking evaporation process, where a Planck mass BH evaporates in a Planck time radiating at the Planck luminosity; this is speculative and beyond current physics.
}
\end{table}
Here we consider the upper bound on the power, or  luminosity, of any process.
In Tab.~\ref{tab:luminosity} we list some of the most powerful and luminous known events. 
Can these values be surpassed by some other, as-yet-unknown process? 
The rate of energy emission for all known interactions increases with acceleration, which in turn increases as bodies get closer together. 
It is therefore not extremely surprising that possible bounds on luminosities are related to maximum-tension principles~\cite{Gibbons:2002iv,Barrow:2014cga}. 
And because BHs represent the limit of short-distance physics, one also gets a natural liaison between bounds on luminosity and gravitation.
In four spacetime dimensions, there is one constant with dimensions of luminosity that can be built from the classical universal constants; this is the Planck luminosity,
\begin{equation}
  {\cal L}_\textrm{P}=\frac{c^5}{G} \approx 3.6\times 10^{52}\,{\rm W}\,.
\end{equation}
This luminosity controls gravitational processes; therefore, it is no wonder that the first LIGO observation of a BH merger was the most powerful astrophysical event ever seen~\cite{2016PhRvL.116f1102A,2016PhRvL.116x1102A}.

\begin{figure}[t]
\centering
\includegraphics[width=0.45\textwidth,clip]{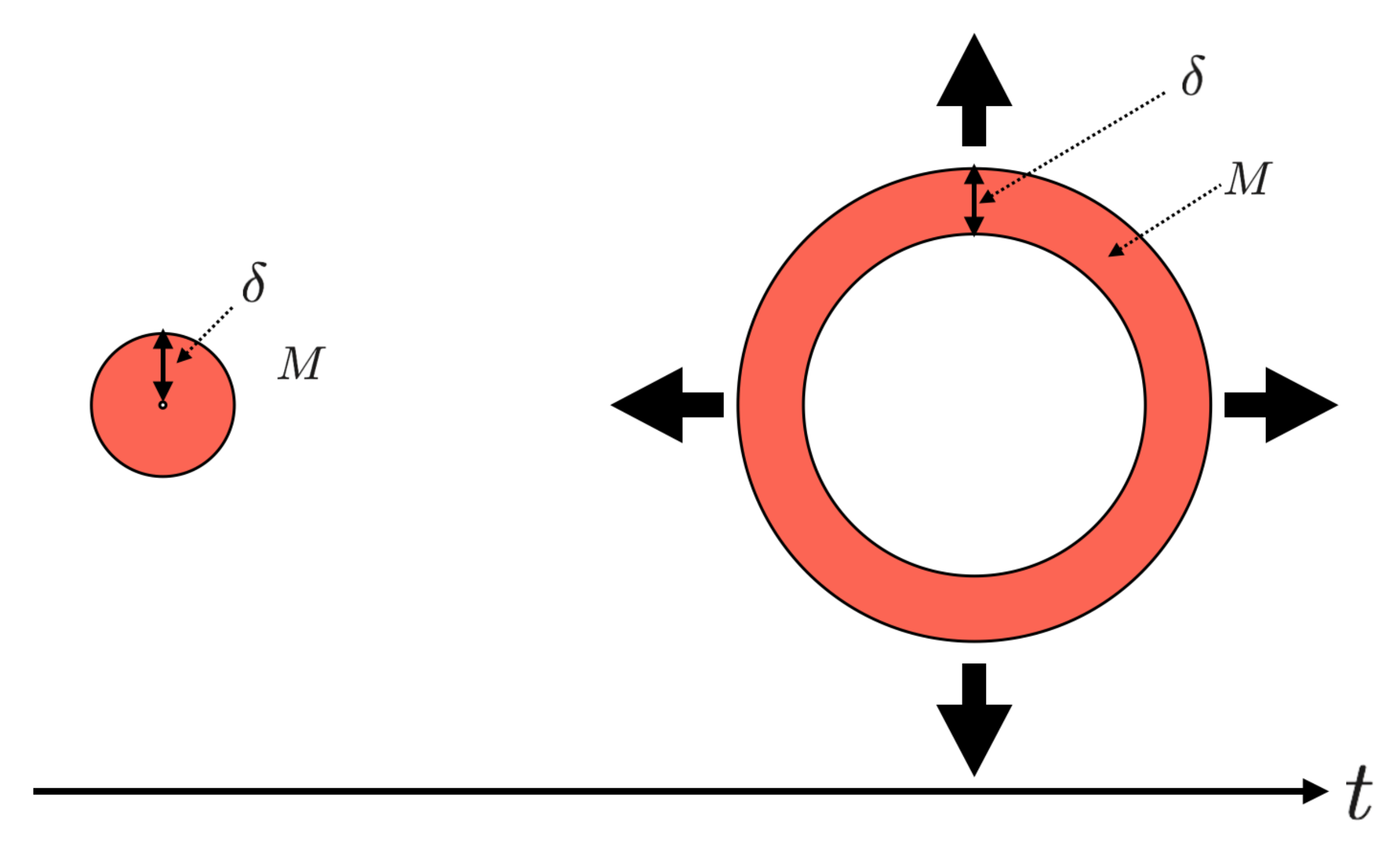}
\caption{\label{Fig_Dyson_Bound} 
An outgoing shell of radiation and its past appearance. 
The luminosity bound ${{\cal L}\lesssim{\cal L}_\textrm{P}/2}$ is imposed by requiring that in the past, when the shell was focussed near the origin, the spacetime does not contain a horizon.
}
\end{figure}

A simple Newtonian thought experiment reveals how BHs regulate the maximum luminosity.
Consider a process with finite duration $\delta t$, which produced some radiation with total energy ${Mc^{2}}$ contained in a shell of thickness ${\delta \! \approx \! c \delta t}$ (see Fig.~\ref{Fig_Dyson_Bound}).
Imagine evolving this system back into the past, so that the shell is focused near the origin where its self-gravity is large. 
In order to have escaped its self-gravity we must impose that a horizon was not present, we require
\begin{equation}
  \frac{2 G M}{c^2 }\lesssim \delta\,.
\end{equation}
The power, or luminosity, is therefore bounded by
\begin{equation}
  {\cal L}\simeq \frac{Mc^{2}}{\delta t}\lesssim\frac{{\cal L}_\textrm{P}}{2}\,.
\end{equation}
It was conjectured by Thorne~\cite{Deruelle:1984hq} (see also \cite{Gibbons:2002iv,Cardoso:2013krh,Barrow:2014cga}) that the Planck luminosity is in fact an upper limit for the luminosity of any process in the Universe.  
\section{Formulating a maximum luminosity conjecture}\label{sec:Formulating a maximum luminosity conjecture}
The above thought experiment suggests that while spacetimes with luminosities above the Planck value may exist, such spacetimes would necessarily contain an horizon, and a singularity in their past history.
To test this we studied a simple system, mimicking as far as possible the setup of Fig.~\ref{Fig_Dyson_Bound}.
We consider spherically symmetric collapse in the full Einstein-Klein-Gordon theory described by the action,
\begin{equation}\label{eq-action}
  S[g_{\mu\nu},\Phi]=\int \textrm{d}^{4}x\; \sqrt{-g}\left(\frac{c^{4}R}{8\pi G}-g^{\mu\nu}\partial_{\mu}\Phi\partial_{\nu}\Phi\right).
\end{equation}
Details of the equations of motion deriving from this action, and the numerical techniques used to evolve them are given in appendix \ref{app:A}.

In this section the numerical evolution of spherically symmetric scalar field shells will be considered, and the resultant luminosities calculated. We studied initial data describing a spherical shell of the scalar field
\begin{align}
  \Phi(t=0,r)&=\Phi_{0}(r)=A\left(\frac{r}{r_{0}}\right)^{7}e^{-\left(\frac{r-r_{0}}{w}\right)^{2}}\,, \label{eq:initial_scalar_pulse_profile} \\
  \Pi(t=0,r)&=\pm\left( \Phi^{\prime}_{0}(r)+\frac{\Phi_{0}(r)}{r}\right)\,. \label{eq:conjmomemtum}
\end{align}
The function $\Phi_{0}(r)$ describes the initial pulse profile ($A$, $w$ and $r_{0}$ are the amplitude, width and radius of the pulse, respectively). 
The function $\Pi$ is the conjugate momentum to the field $\Phi$ (a positive [negative] sign in Eq.~\ref{eq:conjmomemtum} gives an outgoing [ingoing] pulse). 
The initial data for the spacetime metric are also described in appendix \ref{app:A}.

\begin{figure*}[t]
    \centering
    \subfloat[]{{\includegraphics[width=0.458\textwidth]{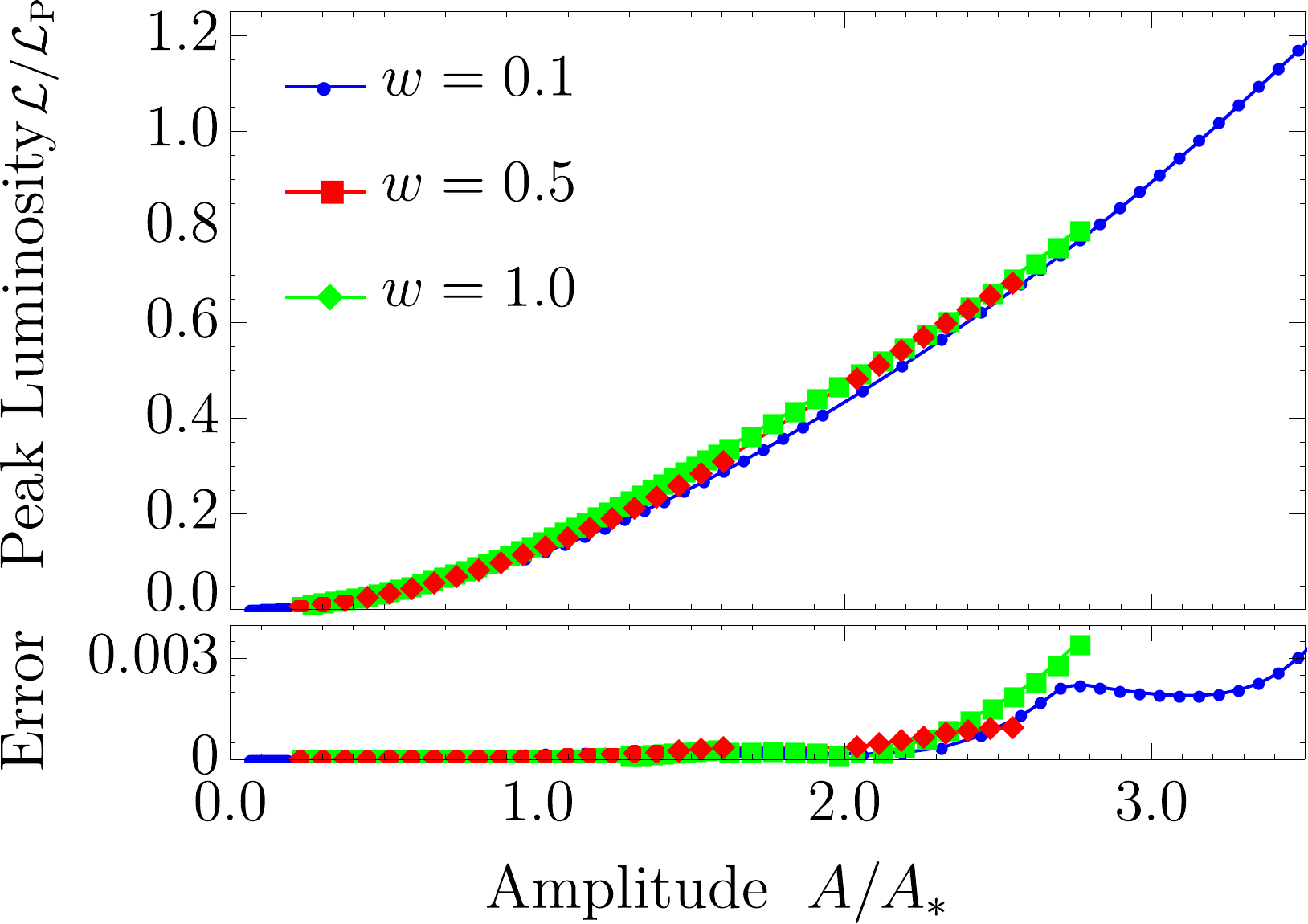} }}%
    \qquad
    \subfloat[]{{\includegraphics[width=0.475\textwidth]{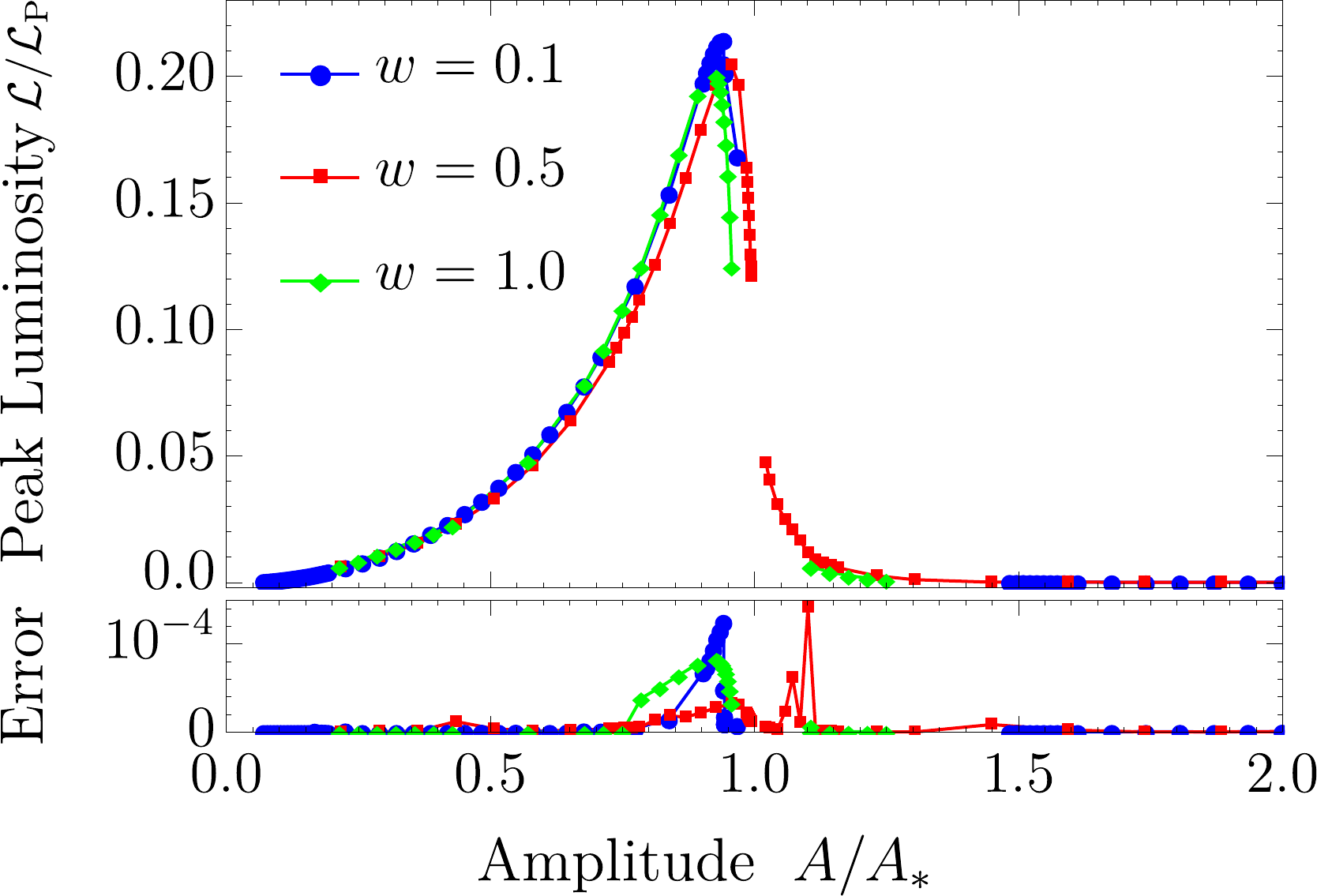} }}%
    \caption{\label{fig:example}
    Panel (a): the peak luminosity measured by an observer at infinity for the spherically symmetric outgoing scalar pulse as function of its amplitude, for 3 different values of the pulse width. 
    The luminosity scales quadratically with the amplitude, and there is no obstacle to obtaining luminosities above ${\cal L}_\textrm{P}$. 
    Panel (b): the peak luminosity for the ingoing field pulse. 
    For small amplitudes the field behaves as in flat spacetime and the peak flux scales quadratically with the amplitude.
    At larger amplitudes, gravitational backreaction becomes important and eventually a horizon forms, suppressing the peak luminosity. 
    The critical amplitude $A_{*}$, is the threshold of horizon formation.
    In both panels the lower plot shows the fraction error on the peak luminosity shown in the upper plot (the calculation of the numerical error is described in appendix \ref{app:A}).
}
\end{figure*}

First, we take initial conditions for an outgoing pulse, evolve the system forward in time and record the peak luminosity, ${\cal L}_{\rm peak}$. 
Our results are summarised in Fig.~\ref{fig:example}(a) where we plot ${\cal L}_{\rm peak}$ as a function of the amplitude for different pulse widths.
Over the range of the amplitudes plotted the peak luminosity ${{\cal L}_{\rm peak} \! \propto \! A^2}$ (as expected for waves in nearly flat spacetimes) and the peak luminosity can {\it exceed the Planck luminosity}. 
The quadratic scaling of the peak luminosity with $A$ holds up until the point where the mass energy in the shell becomes comparable to $\sim r_{0}/2$, at which point a horizon forms trapping the outgoing shell.
However, with a suitable choice of the initial radius, $r_{0}$, amplitude, $A$, and width, $w$, parameters for the shell there is no upper bound to the luminosity that can be achieved using the initial data of the form in Eqs.~\ref{eq:initial_scalar_pulse_profile} and \ref{eq:conjmomemtum}.
This provides a concrete counter-example to any naive bound on the maximum luminosity. 

However, there remains the question of producing such initial data from reasonable conditions arising in the Universe. 
To investigate this we take initial conditions for the ingoing pulse, and evolve this system forwards in time.
We find that for amplitudes larger than some critical value, $A_{*}$, the scalar field pulse shrinks inside a horizon and forms a singularity.
Therefore, the maximal development of the initial data for an outgoing pulse with an amplitude ${A\!>\!A_{*}}$ necessarily contains a past horizon and past singularity.
Our results for the ingoing pulse (or time-reverse outgoing pulse) are summarised in Fig.~\ref{fig:example}(b).
For small amplitudes we have ${{\cal L}_\textrm{P} \! \propto \! A^2}$; the self-gravity of the pulse is weak, so it passes through itself at the origin without significant interaction and goes back out to infinity.
However, the peak luminosity reaches a maximum value ${\approx \! 0.2 {\cal L}_\textrm{P}}$ (the peak value depending only weakly on the width). 
For initial amplitudes above the critical value, ${A \! \geq \! A_{*}}$ \cite{PhysRevLett.70.9}, the ingoing pulse forms a horizon which retains a fraction of the energy, and the luminosity at infinity is reduced. 
The peak luminosities obtained here are the largest obtained to date in numerical simulations, but are still below the Planck value.
One might try to achieve still larger luminosities by placing a ``rigid'' sphere at some radius $R$ to reflect the incoming shell and prevent it from forming a BH at the origin. However, such a sphere would need to support pressures of ${\sim \! Mc^{2}/(4\pi R^{2}w)}$, corresponding to a tension ${\sim \! Mc^{2}/w}$. If the maximum tension principle holds \cite{Gibbons:2002iv,Barrow:2014cga}, then such shell can only work below the Planck luminosity.
These numerical results lead us to make the following conjecture.

{\em Maximum Luminosity Conjecture:} the luminosity of any process in a spacetime satisfying the maximum tension principle and free from past horizons is bounded above by the Planck luminosity.

The luminosity is to be understood as the rate of change of energy (specifically the Misner-Sharp mass \cite{Misner:1964je}, or energy, see appendix \ref{app:A}) with proper time at future null infinity in an asymptotically flat spacetime.

\section{Testing the maximum luminosity conjecture}
Other attempts at exceeding the Planck luminosity using other fundamental fields seem to stumble upon BHs as well.
Consider the electromagnetic and gravitational radiation emitted during the head-on collision of two charged bodies infalling from rest,
of equal mass $M/2$ and opposite charge $\pm Q$. 
The classical Larmor formula for electromagnetic (EM) waves and the quadrupole formula for gravitational wave (GW) radiation yields \cite{Zilhao:2013nda}, 
\begin{align}
  {\cal L}_\textrm{GW}(z)=\frac{{\cal B}^3}{480} \left(\frac{r_G}{z}\right)^5\,,\;\;
  {\cal L}_\textrm{EM}(z)= \frac{{\cal B}^2}{192}\frac{r_e}{z}\left(\frac{r_G}{z}\right)^3\,,
\end{align}
where ${{\cal B} \! = \! 1 \! + \!Q^2/(4\pi\epsilon_{0}GM^2c^2)}$, the ``classical electric radius'' is ${r_e \! = \! 2Q^2/(4\pi\epsilon_{0}Mc^2)}$, the ``gravitational radius'' is ${r_G \! = \! GM/c^2}$, and $z$ is the distance between the two objects. 
If the particle is outside its own Schwarzschild radius ${z \! > \! r_e \! > \! r_G}$ then the total luminosity satisfies ${{\cal L}_\textrm{EM}  +  {\cal L}_\textrm{GW} \! <  \! {\cal L}_\textrm{P}}$.
Full nonlinear simulations support this conclusion~\cite{Zilhao:2013nda,zilhaoprivate} and the result can be generalised to arbitrary, non-relativistic motions.

The above calculation considered only two point-like, structureless particles. 
It is conceivable that there exist composite systems which emit above the conjectured bound. 
However, finding such a system is challenging. A simple setup which seemingly provides higher luminosities is the straightforward combination of $N$ separate binaries, each radiating at a fixed luminosity: if the luminosities add, then a sufficiently large number of binaries should violate the conjectured bound. To be specific: the first LIGO event, GW150914, consisted of two comparable BHs, each $\sim\! 30M_{\odot}$. When they entered band, they were separated by $\sim\! 300 \,{\rm km}$ and radiating gravitational waves at a luminosity of $\sim\! 10^{-3} {\cal L}_\textrm{P}$~\cite{Abbott:2016bqf,Keitel:2016krm}. Can one take $10^3$ or more GW150914-like binaries each separated by, say, $10^5 \,{\rm km}$, such that the total luminosity adds up to more than ${\cal L}_\textrm{P}$? However, to guarantee that the luminosities do indeed add, one must ensure that each binary is stable and radiating for at least the light travel time between the binaries, so that the flux crossing a sphere at large radii is the same at each point on that sphere.
Consider an equal mass binary with total mass $M$ and separation $R$.
From the quadrupole formula \cite{Misner_1973} the radiated GW power depends only on the dimensionless separation ${x \! \equiv \! c^{2}R/GM}$,
\begin{equation}
  P_{\textrm{GW}}\left(x\right) = \frac{2}{5}{\cal L}_\textrm{P} x^{-5} \,,
\end{equation}
and the binary radiates for a total lifetime \cite{Misner_1973}
\begin{equation}
  \tau\left(x\right) = \frac{5}{64}\frac{GM}{c^{3}} x^{4} \,. 
\end{equation}
We now wish to double the GW luminosity by adding together two such systems (interference between the GWs from the two systems is neglected). At what distance apart, $d$, should the two binaries be placed? 
The distance must be sufficiently large that the two binaries remain radiating on quasi-circular orbits without disturbing one another over a timescale of at least one orbital period; this minimum distance is estimated here as the GW wavelength (which can be calculated from Kepler's law). As discussed above, the distance must also be less than the light travel distance of the evolution time. Therefore,
\begin{align}
  &d_{\textrm{min}}\left(x\right) = \frac{c}{\omega} = \frac{GM}{2c^{2}}x^{3/2} \,,\\ &d_{\textrm{max}}\left(x\right) = c\tau(x) = \frac{5}{64}\frac{GM}{c^{2}} x^{4}  \,.
\end{align}
Other estimates for the the minimum distance can be explored; for example, the minimum distance can be estimated as twice the orbital diameter, $d_{\textrm{min}}=4R$. We have verified that the results of the following calculation do not depend sensitively on the choice of $d_{\textrm{min}}$.
We now wish to further increase the GW luminosity by adding together many such systems; what is the maximum achievable luminosity? 
Packing $N$ binaries into a cube of side length $d_{\textrm{max}}$, it would appear that the limit is ${N \! < \! (d_{\textrm{max}}/d_{\textrm{min}})^{3}}$. 
However, we must also ensure that the total mass in the cube is sufficiently low that we avoid forming a single BH trapping all of the radiation; the hoop conjecture~\cite{Thorne:1972,2009arXiv0903.1580G} forces us to have ${2GNM/c^{2}\! < \! d_{\textrm{max}}}$. 
Therefore, the maximum number of binaries possible while still having the total luminosity scale linearly is
\begin{align} N(x) = &\;\textrm{min}\left( \left(\frac{d_{\textrm{max}}}{d_{\textrm{min}}}\right)^{3} , \frac{c^{2}d_{\textrm{max}}}{2GM} \right) \\ N(x)= &\;\textrm{min}\left( \frac{5^{3}}{2^{15}}x^{15/2} , \frac{5}{2^{7}}x^{4} \right) \,. 
\end{align}
The GW luminosity of the ensemble of BHs is
\begin{equation}
{\cal L}\left(x\right) \!=\! N(x) P_{\textrm{GW}}(x) \!=\! \textrm{min}\left( \frac{5^{2}}{2^{14}}x^{5/2} , \frac{1}{2^{6}}x^{-1} \right){\cal L}_\textrm{P} ,
\end{equation}
which has a maximum value of ${{{\cal L}_{\textrm{max}} \! \approx \! 8.0 \! \times \! 10^{-3}{\cal L}_\textrm{P}}}$ at ${x \! \approx \! 1.9}$, satisfying the conjectured bound.
Again, even in this quasi-Newtonian analysis we see the role that BH horizons (or at least their absence) play in regulating the luminosity and enforcing the conjectured bound.
\section{Going beyond the maximum luminosity?}
The conjectured bound on the maximum possible luminosity appears fairly secure, it has withstood all our attempts thus far to falsify it. We conclude by presenting an analytic example pointing to a possible failure of the conjecture.

The thought experiment illustrated in Fig.~\ref{Fig_Dyson_Bound} motivated the upper luminosity bound by considering a time-reversed outgoing shell forming a past horizon. 
However, absorption may, in principle, help avoid the formation of a past horizon and open up the possibility of exceeding the conjectured bound.
One can imagine an extended matter distribution which gradually absorbs the energy as it travels inwards. If the attenuation is sufficiently strong it may reduce the energy in the shell to a point where no horizon forms. We consider this here. 

Consider a static, spherical ``star'' with a typical size $a$ and mass $M$, with a spacetime metric 
\begin{align}\label{eq:a}
  \textrm{d} s^{2} = &-\bigg(1-\frac{2Gm(r)}{rc^{2}}\bigg) c^{2} \textrm{d} t^{2}+ \textrm{d} r^{2}\left(1-\frac{2Gm(r)}{rc^{2}}\right)^{-1} \nonumber \\& + r^{2} \left(\textrm{d} \theta^{2}+\sin^{2}\theta \textrm{d} \phi^{2}\right) \,, \\
  &\quad\quad\textrm{with }\, m(r) = M r^{3}\left(a^2+r^2\right)^{-3/2} \,.
\end{align}
The stress-energy tensor of the matter in the star may be calculated from the Einstein equations, ${T_{\mu\nu}\!=\!(c^{4}/8\pi G){\cal G}_{\mu\nu}}$. 
Details are given in appendix \ref{app:B}, here we simply note that the matter is physically reasonable and satisfies the null, weak, and dominant energy conditions \cite{Hawking:1973uf}.

\begin{figure*}[t]
    \centering
    \subfloat[]{{\includegraphics[width=0.2\textwidth,valign=c]{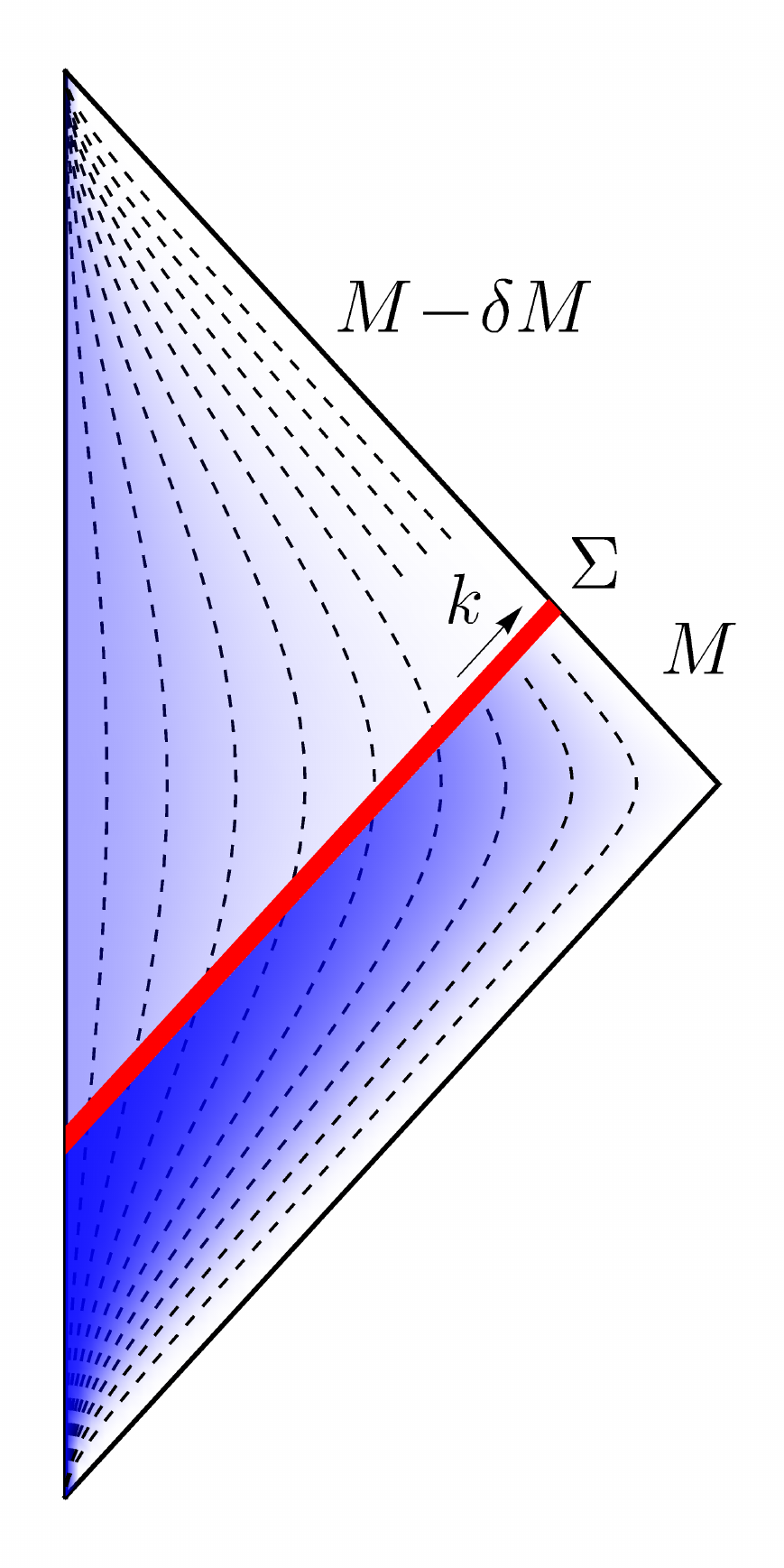} }}%
    \qquad
    \subfloat[]{{\includegraphics[width=0.5\textwidth,valign=c]{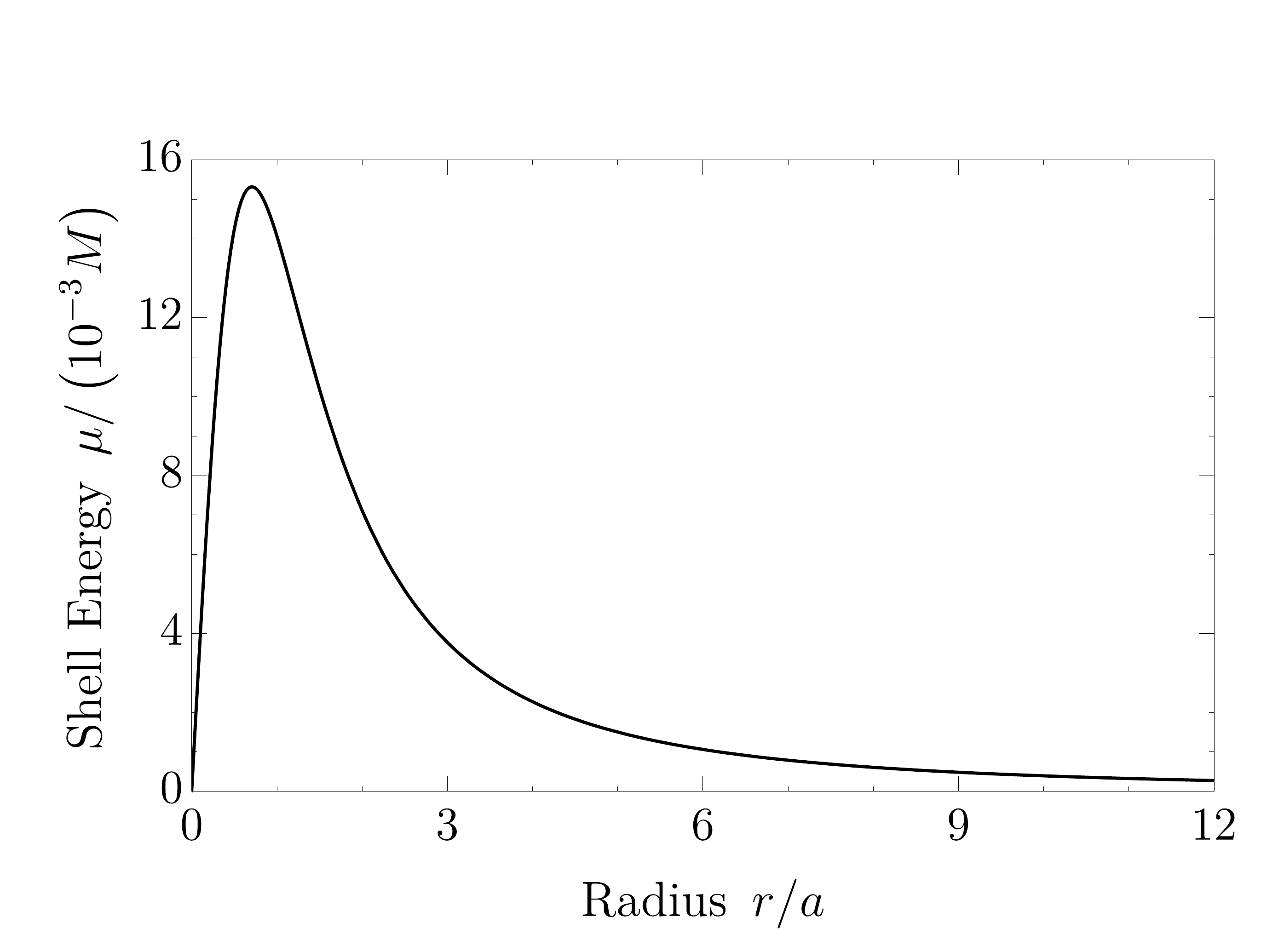} }}%
    \caption{\label{fig:Plummer}
    Panel (a): the conformal diagram (or Penrose diagram) for the spacetime described in the text. The blue shading indicates the density of matter in the star and the red line indicates the outgoing spherical shell. In the past of the shell the spacetime is asymptotically Schwarzschild with mass $M_{-}$, and in the future the spacetime is asymptotically Schwarzschild with mass ${M_{+} \! = \! M_{-} \! - \! \delta M}$. The difference, $\delta M$, is radiated out to infinity along $\Sigma$.
    Panel (b): the energy density in the the outgoing spherical shell as a function of radius. Near the origin the density of the star is approximately constant, and as the outgoing pulse passes through the star each layer adds more energy to the pulse, and $\mu$ increases linearly. At large radii the spacetime is vacuum, and the energy density in the shell decreases as $r^{-2}$ as the shell spreads out in space.
}
\end{figure*}

Consider two copies of this spacetime with masses $M$ and ${M\!-\!\delta M}$. We cut these spacetimes along the surface of an outgoing null geodesic cone, $\Sigma$, with its vertex at the origin. The two spacetimes may be joined along $\Sigma$, placing the spacetime with the lower mass in the future. Care must be taken to ensure the appropriate Israel boundary conditions are satisfied at the junction \cite{Israel1966,2002gr.qc.....7101P}. The resulting spacetime satisfies the Einstein equations, with some additional matter on $\Sigma$; the stress-energy tensor of this additional matter may be found by considering the geometry of $\Sigma$ (see appendix \ref{app:B}) and we find that it is that of ``null dust'', i.e.\
\begin{align} \label{eq:delta}
  T^{\mu\nu}_{\Sigma} &= (-k_{\rho} u^{\rho})^{-1} \delta(\tau-\tau_{0}) \,\mu(r)\, k^{\mu}k^{\nu}\,,\\ &\textrm{with }\;\mu(r) =\; \frac{r\delta M}{16\pi\left(a^{2}+r^{2}\right)^{3/2}}\,,
\end{align}
$k^{\mu}$ is an outward future-pointing null vector (see Fig.~\ref{fig:Plummer} (a)), $u^{\mu}$ is the four-velocity of an observer crossing $\Sigma$, and ${\delta(\tau-\tau_{0})}$ is a $\delta$-function in the observer's proper time peaked on $\Sigma$.

An observer at future null infinity will measure an unboundedly large luminosity from this system due to the $\delta$-function in Eq.~\ref{eq:delta}. 
Such a spacetime, while admittedly being somewhat contrived, could be practically realised, by a sufficiently advanced civilisation, as the limit of the following construction. 
First they must construct a spherical ball of matter in an excited state. 
Then they must arrange for a small piece, or ``trigger region'', of the matter in the centre to drop into a stable state, releasing its energy as a small burst of radiation. 
As this burst travels out through the ball it can stimulate the outer layers to drop into the stable state adding their energy to the outgoing null shell; in this way the luminosity of the burst increases as is travels out through the ball (see Fig.~\ref{fig:Plummer}(b)). 
Think of a spherical laser, with the ball playing the role of the lasing medium. 
This system can release a large amount of energy (the amount being set by the size of the ball) in a short amount of time (the duration being set by the size of the ``trigger region'') with an arbitrarily large luminosity.

This example illustrates a possible way to violate the maximum luminosity conjecture.
The underling mechanism is that the radiation must arise from stimulated emission along an outgoing null cone in an extended object, or ``star''; in this way a finite energy can be radiated in an arbitrarily thin shell with a super-Planckian luminosity.
In the time reversed picture the radiation is gradually attenuated by the extended ``star'' as in propagates inwards, and the formation of a past horizon is avoided.
By itself, we do not consider the spacetime in Eq.~\ref{eq:a} to be a convincing counterexample to the conjecture. 
Because of the way the spacetime was constructed we do not have access to the equations of motion for the matter in the ``star'', this means that the nature of the matter remains rather unclear (although it does satisfy the energy conditions, see appendix \ref{app:B}), and questions concerning the star's stability cannot be addressed. 
However, we hope that this example will serve as a prototype for future efforts to probe the maximum luminosities attainable in this way.

\section{Discussion}
Bounds on physical quantities have played a pivotal role in the development of science.
All experimental observations so far are consistent with there being a maximum possible luminosity, or power, ${{\cal L}_\textrm{P}\!=\!c^{5}/G}$.
The majority of our attempts to exceed this bound either failed, had horizons in their past, or were in conflict with maximum tension principles~\cite{Gibbons:2002iv,Barrow:2014cga}.
Event horizons play a key role in regulating this maximum luminosity.

The numerical simulations of the Einstein-Klein-Gordon system performed here are also consistent with the conjectured bound.
These simulations provide the clearest evidence yet that any attempt to make a precise mathematical formulation of the maximum luminosity conjecture must make reference to the non-existence of past horizons. 
These simulations have also yielded the largest luminosities yet seen in numerical relativity.

Finally, we have presented a prototype for possible future attempts to exceed the conjectured bound. 
The key physical observation is that the emission should come from an extended object, and be triggered by an outwardly propagating null signal. 
Only in this way can a finite energy be radiated in an arbitrarily thin shell without the energy being trapped inside of a horizon. 

We hope that this work will help motivate further investigations of the maximum possible power and the role of the Planck luminosity, $\mathcal{L}_{\textrm{P}}\!=\!c^{5}/G$.

\begin{acknowledgments}
We are grateful to Kinki University in Osaka and Nagoya University for kind hospitality while part of this work was done.
We thank Hirotaka Yoshino, Bruce Allen, Miguel Zilh\~ao and Tomohiro Harada for useful discussions regarding this work.
V. C. and C. M. acknowledges financial support provided under the European Union's H2020 ERC Consolidator Grant ``Matter and strong-field gravity: New frontiers in Einstein's theory'' grant agreement no. MaGRaTh--646597. Research at Perimeter Institute is supported by the Government of Canada through Industry Canada and by the Province of Ontario through the Ministry of Economic Development $\&$
Innovation.
This project has received funding from the European Union's Horizon 2020 research and innovation programme under the Marie Sklodowska-Curie grant agreement No 690904.
This work was supported by JSPS KAKENHI Grant Numbers JP16K17688, JP16H01097 (CY), and KMI visitor program.
This article is based upon work from COST Action CA16104 ``GWverse'', supported by COST (European Cooperation in Science and Technology).
\end{acknowledgments}

\bibliographystyle{apsrev4-1}
\bibliography{refs}

\clearpage

\appendix

\section{Numerical evolution of the Einstein-Klein-Gordon system}\label{app:A}
The equations of motion deriving from the Einstein-Klein-Gordon action in Eq.~\ref{eq-action} are
\begin{align} \label{eq:fieldeqns}
  {\cal G}_{\mu\nu} & = 8\pi T_{\mu\nu}\, , \,\textrm{ and }\; \Box\Phi = 0 \, ,
\end{align}
where $\Box$ is the D'Alembertian operator, ${\cal G}_{\mu\nu}$ is the Einstein tensor associated with the metric $g_{\mu\nu}$, and
$T_{\mu\nu}$ is the energy momentum tensor of the massless scalar field, given by
\begin{equation}
T_{\mu\nu}=-\frac{1}{2}g_{\mu\nu}(\nabla\Phi)^{2}+\nabla_{\mu}\Phi\nabla_{\nu}\Phi \, .
\end{equation}
In this section we use natural units where $G=c=1$.
We evolve these equations using the G-BSSN formulation \cite{Brown:2009dd,Alcubierre:2010is}, which we briefly summarize here.
The G-BSSN formulation is a generalization of the BSSN formulation \cite{Shibata:1995we,Baumgarte:1998te}, which is in turn based on the ADM formalism \cite{Arnowitt:1962hi}, to the case of curvilinear coordinates.
The line element in ADM formulation is written as
\begin{equation}
  \textrm{d}s^{2} = -\alpha^{2}\textrm{d}t^{2}+\gamma_{ij}(\textrm{d}x^{i}+\beta^{i}\textrm{d}t)(\textrm{d}x^{j}+\beta^{j}\textrm{d}t)\,.
\end{equation}
In the G-BSSN formulation, the spatial metric $\gamma_{ij}$ and the extrinsic curvature $K_{ij}$ are decomposed as
\begin{equation}
  \gamma_{ij}=e^{4\phi}\tilde{\gamma}_{ij}\;\;\textrm{and }\;\;  K_{ij}=e^{4\phi}\tilde{A}_{ij}+\frac{1}{3}\gamma_{ij}K\,, \label{conformal decomposition of the spatial metric}
\end{equation}
where $K=\gamma^{ij}K_{ij}$ and $\tilde{\gamma}\equiv\mbox{det}(\tilde{\gamma}_{ij})$ is determined from the Lagrangian evolution equation, $\partial_{t}\tilde{\gamma}=0$.
Furthermore, we introduce a reference metric $\bar{\gamma}_{ij}$, and an auxiliary vector field 
\begin{equation}
  \tilde{\Lambda}^{k}=\tilde{\gamma}^{ij}(\tilde{\Gamma}^{k}_{ij}-\bar{\Gamma}^{k}_{ij})\,,
\end{equation}
where $\Gamma^{k}_{ij}$ $\left[\bar{\Gamma}^{k}_{ij}\right]$ are the Christoffel symbols associated with $\gamma_{ij}$ $\left[\bar{\gamma}_{ij}\right]$.

We now specialise to the case of spherical symmetry, where we have
\begin{align}
  \tilde{\gamma}_{ij} &= \mbox{diag}(a,br^{2},br^{2}\sin^{2}\theta)\,,\\
  \bar{\gamma}_{ij} &= \mbox{diag}(1,r^{2},r^{2}\sin^{2}\theta)\,, \\
  \tilde{A}_{ij} &= \mbox{diag}(A,Br^{2},Br^{2}\sin^{2}\theta)\,, \\
  \beta^{i}&=(\beta,0,0)\,,\\
  \tilde{\Lambda}^{i}&=(\tilde{\Lambda},0,0)\,.
\end{align}
The evolution equations are given by
\begin{align}
  \partial_{t}\phi=&\beta\phi^{\prime}-\frac{1}{6}\alpha K+ \frac{1}{6}\mathcal{B}\,,\\
  \partial_{t}a=&\beta a^{\prime}+2a\beta^{\prime}-2\alpha A - \frac{2}{3}a\mathcal{B}\,,\\
  \partial_{t}b=&\beta b^{\prime}+2\beta\frac{b}{r}-2\alpha B - \frac{2}{3}b\mathcal{B}\,,\\
  \partial_{t}K=&\beta K^{\prime}-\mathcal{D}+\alpha\left(\frac{1}{3}K^{2}+\frac{A^{2}}{a^{2}}+2\frac{B^{2}}{b^{2}}\right)\nonumber\\&+4\pi\alpha\left(E+S\right)\,,\\
  \partial_{t}A=&\beta A^{\prime}+2A\beta^{\prime}\nonumber\\&+e^{-4\phi}\left(-\mathcal{D}_{rr}^{\textrm{TF}}+\alpha\left(R_{rr}^{\textrm{TF}}-\frac{16\pi}{3}\Phi^{\prime 2}\right)\right)\nonumber\\&+\alpha \left(KA-2\frac{A^{2}}{a}\right)- \frac{2}{3}A\mathcal{B}\,,\\
  \partial_{t}B=&\beta B^{\prime}+\frac{e^{-4\phi}}{r^{2}}\left(-\mathcal{D}_{\theta\theta}^{\textrm{TF}}+\alpha\left(R_{\theta\theta}^{\textrm{TF}}+\frac{8\pi br^{2}}{3a}\Phi^{\prime 2}\right)\right)\nonumber\\&+\alpha\left(KB-2\frac{B^{2}}{b}\right)+2\frac{\beta}{r}B- \frac{2}{3}B\mathcal{B}\,,\\
  \partial_{t}\tilde{\Lambda}=&\beta\tilde{\Lambda}^{\prime}-\beta^{\prime}\tilde{\Lambda}\nonumber\\&+\frac{2\alpha}{a}\left(\frac{6 A\phi^{\prime}}{a}-\frac{2}{3}K^{\prime}+8\pi\frac{\Pi\Phi^{\prime}}{e^{2\phi}\sqrt{a}}\right)\nonumber\\&+\frac{\alpha}{a}\left(\frac{a^{\prime}A}{a^{2}}-\frac{2b^{\prime}B}{b^{2}}+4B\frac{a-b}{rb^{2}}\right)\nonumber\\&+ \left(\frac{2}{3}\tilde{\Lambda}\mathcal{B}+\frac{\mathcal{B}^{\prime}}{3a}\right)+\frac{2}{rb}\left(\beta^{\prime}-\frac{\beta}{r}\right)\nonumber\\&-2\frac{\alpha^{\prime}A}{a^{2}}+\frac{1}{a}\beta^{\prime\prime}.
\end{align}
In the above evolution equations a prime denoted a $r$-derivative, a superscript TF denotes the trace-free part, and we have defined $\mathcal{D}_{ij}\equiv D_{i}D_{j}\alpha$, $\mathcal{D}\equiv\gamma^{ij}D_{i}D_{j}\alpha$, $\mathcal{B}\equiv \tilde{D}_{k}\beta^{k}$.
The Hamiltonian and momentum constraint are given by
\begin{align}
  H \equiv&\left(\frac{\phi^{\prime\prime}}{a}+\frac{\phi^{\prime 2}}{a}-\left(\frac{a^{\prime}}{2a^{2}}-\frac{b^{\prime}}{ab}-\frac{2}{ar}\right)\phi^{\prime}\right)\nonumber\\&-\frac{\tilde{R}}{8}+\frac{e^{4\phi}}{8}\left(\frac{A^{2}}{a^{2}}+2\frac{B^{2}}{b^{2}}\right)\nonumber\\ &-\frac{e^{4\phi}}{12}K^{2}+\pi \left( \frac{\Pi^{2}+\Phi^{\prime 2}}{a} \right)=0\,,\label{Eq:Hamiltonian constraint}\\
  M\equiv&\,6\phi^{\prime}\frac{A}{a}+\frac{A^{\prime}}{a}-\frac{a^{\prime}A}{a^{2}}+\frac{b^{\prime}}{b}\left(\frac{A}{a}-\frac{B}{b}\right)+\frac{2}{r}\left(\frac{A}{a}-\frac{B}{b}\right)\nonumber\\ &-\frac{2}{3}K^{\prime} +\frac{8\pi}{e^{2\phi}\sqrt{a}}\Pi\Phi^{\prime}=0\,.\label{Eq:momentum constraint}
\end{align}
Finally, the equations of motion for the scalar field are given by
\begin{align}
  \dot{\Pi}=&\beta\Pi^{\prime}+\Big(\frac{2}{3}\alpha K+2\alpha\frac{B}{b}+\beta^{\prime}\Big)\Pi\nonumber\\
&+\Big(\alpha^{\prime}e^{-2\phi}a^{-1/2}+2\alpha\phi^{\prime}e^{-2\phi}a^{-1/2}-\frac{1}{2}\alpha e^{-2\phi}a^{-3/2}a^{\prime}\nonumber\\
&+\alpha e^{-2\phi}a^{-1/2}\frac{b^{\prime}}{b}+2\alpha e^{-2\phi}a^{-1/2}\frac{1}{r}\Big)\Phi^{\prime}\nonumber\\&+\alpha e^{-2\phi}a^{-1/2}\Phi^{\prime\prime}\,,\\
\dot{\Phi}=&\beta\Phi^{\prime}+\frac{\alpha}{e^{2\phi}\sqrt{a}}\Pi\,.
\end{align}
In this study, we solve these equations using a partially implicit Runge-Kutta method.\cite{Baumgarte:2012xy} 
The gauge condition for the lapse function and the shift vector are the harmonic gauge and the gamma driver respectively.

We wish to evaluate the luminosity, which is the rate of change of an energy, or mass. 
In spherical symmetry the existence of the Kodama vector, $K^{\mu}$, \cite{Kodama:1979vn} allows for the definition of a flux vector $S^{\mu}\equiv T^{\mu}_{\nu}K^{\nu}$ satisfying the following conservation law,
\begin{equation}\label{eq:conservation law}
  \partial_{\mu}(\sqrt{-g}S^{\mu})=0\,.
\end{equation}
From this conservation law we have the quasi-local conserved \emph{Misner--Sharp} mass \cite{Misner:1964je},
\begin{equation}
  M(t,r_{0})=\int_{r<r_{0}} \textrm{d}x^{3}\;S^{t}\alpha\sqrt{\gamma}\,,
\end{equation}
and the associated luminosity
\begin{align}\label{eq:integrate conservation law}
\frac{\partial}{\partial t}&M(t,r_{0})=-\mathcal{L}(t,r_{0})\,,  \\ \textrm{with }\;   &\mathcal{L}(r)=4\pi\alpha e^{6\phi}a^{1/2}br^{2}S^{r}\,.
\end{align}

We evolve the initial data described in the main text according the above formalism, and record the peak luminosity $\mathcal{L}_{\rm peak}$ at four different radii: $2.5r_{0}$, $5.0r_{0}$, $7.5r_{0}$, and $10.0r_{0}$.
To obtain the peak luminosity at infinity, $\mathcal{L}_{\rm peak}(\infty)$, we fit these luminosities with 
\begin{equation}
  {\cal L}_{\rm{peak}}(r)={\cal L}_{\rm{peak}}(\infty)+\frac{c_{1}}{r}+\frac{c_{2}}{r^{2}}\,.\label{asymptotic form}
\end{equation}
The coefficients $c_{1}$, $c_{2}$ and ${\cal L}_{\rm{peak}}(\infty)$ are determined from the measured luminosities at $2.5r_{0}$, $5.0r_{0}$, and $7.5r_{0}$.
We then estimate the error in value obtained for $\mathcal{L}_{\rm peak}(\infty)$ using the difference between the measured luminosities at $10.0r_{0}$ and the value given by Eq.~\ref{asymptotic form}.

\subsection{Initial data}
We now outline the construction of the initial data used to perform the numerical simulations described in Sec.~\ref{sec:Formulating a maximum luminosity conjecture}.
The initial data describes a outgoing or ingoing spherical shell of scalar field, of the form in Eqs.~\ref{eq:initial_scalar_pulse_profile} and \ref{eq:conjmomemtum}.
For simplicity, the metric ansatz of the initial data is conformally flat (that is $a(t=0,r)=1$ and $b(t=0,r)=1$) and $K$ vanishes at the initial time.
Since $\tilde{A}_{ij}$ is traceless, we have $B=-\frac{b}{2a}A$.
$\phi$ and $A$ are given as solutions of the Hamiltonian constraint in Eq.\ref{Eq:Hamiltonian constraint}, and momentum constraint in Eq.~\ref{Eq:momentum constraint}, respectively.
The boundary conditions at the origin are 
\begin{eqnarray}
\phi^{\prime}(t=0,r=0)&=&0,\label{Eq:boundary condition at the origin for phi}\\
A(t=0,r=0)&=&0.\label{Eq:boundary condition at the origin for A}
\end{eqnarray}
Since the spacetime is asymptotic flat, the boundary condition at the outer boundary are
\begin{eqnarray}
\phi(t=0,r\to \infty)&\to&0,\label{Eq:boundary condition at far boundary for phi}\\
A(t=0,r\to\infty)&\to&0.\label{Eq:boundary condition at far boundary for A}
\end{eqnarray}
We solve the Hamiltonian constraint and the momentum constraint for $\phi$ and $A$ under the boundary conditions using a ``shooting method''.
If the condition Eq.\ref{Eq:boundary condition at far boundary for phi} is satisfied, then condition Eq.\ref{Eq:boundary condition at far boundary for A} is also automatically satisfied.
Therefore, the shooting parameter is the value of the scalar field at the origin.

After solving the constraint equations, the initial profile of the lapse function $\alpha$ and the shift vector $\beta$ are determined. 
The profile of the lapse function is given from the condition that the time derivative of $K$ vanishes for the initial data; that is, the lapse function at the initial data is a solution of
\begin{align}
\alpha^{\prime\prime}=&\,-\left(-\frac{a^{\prime}}{2a}+\frac{b^{\prime}}{b}+2\phi^{\prime}+\frac{2}{r}\right)\alpha^{\prime}\nonumber\\
&+ae^{4\phi}\Big\{\beta K^{\prime}+\alpha\left(\frac{1}{3}K^{2}+\frac{A^{2}}{a^{2}}+2\frac{B^{2}}{b^{2}}\right)\nonumber \\ &\quad\quad\quad\;+4\pi\alpha(E+S)\Big\}.
\end{align}
For simplicity, the shift vector vanishes for the initial data.

\section{The geometry of the outgoing null cone} \label{app:B}
In this section we describe the geometry of the outgoing null cone forming the junction between two spacetimes, and the determination of the resultant stress-energy tensor on the cone.
The relevant boundary conditions were first derived in \cite{Israel1966}, here we will use the notation and conventions of the reformulation in \cite{2002gr.qc.....7101P}.
In this section we will work with natural units with $G=c=1$.

Consider the static, spherically symmetric metric given by
\begin{equation}\label{eq:supp_metric}
  \textrm{d} s^{2} = -A(r)^{2} \textrm{d} t^{2} + B(r)^{2}\textrm{d} r^{2} + r^{2} \left( \textrm{d} \theta^{2} + \sin^{2}\theta \, \textrm{d}\phi^{2} \right)\,,
\end{equation}
expressed in spherical coordinates $x^{\mu}=(t,r,\Theta^{A})$, where $\Theta^{A}=(\theta,\phi)$.
Consider the null hypersurface, $\Sigma$, formed by a radial outgoing null geodesic cone. 
The outgoing null geodesic equation may be read off directly from equation \ref{eq:supp_metric}.
The coordinates of $\Sigma$ are $x^{\mu}(y^{a})=(t(r),r,\Theta^{A})$, with $t(r)=\int_{0}^{r}\textrm{d} \bar{r} \, (B(\bar{r})/A(\bar{r}))$, and we use $y^{a}=(r,\Theta^{A})$ as coordinates on $\Sigma$.

The three tangent vectors to $\Sigma$ are given by
\begin{align} 
  e^{\alpha}_{r}=k^{\alpha} &\equiv \frac{\partial x^{\mu}(y^{a})}{\partial r} = \left(B(r)/A(r) , 1,0,0\right), \\
  e^{\alpha}_{\theta} &\equiv \frac{\partial x^{\mu}(y^{a})}{\partial \theta} = \left(0,0,1,0\right), \\
  e^{\alpha}_{\phi} &\equiv \frac{\partial x^{\mu}(y^{a})}{\partial \phi} = \left(0,0,0,1\right),
\end{align}
and satisfy $k_{\alpha}k^{\alpha}=0=e_{A}^{\alpha}k_{\alpha}$. The intrinsic metric on $\Sigma$ is 2 dimensional and is given by the remaining inner products,
\begin{equation} 
  \sigma_{AB}(r,\theta,\phi) \equiv g_{\alpha\beta}e^{\alpha}_{A}e^{\beta}_{B}=\begin{pmatrix} r^{2} & 0 \\ 0 & r^{2}\sin^{2}\theta \end{pmatrix}.
\end{equation}
The normal vector to $\Sigma$ satisfying $N_{\alpha}N^{\alpha}\!=\!N_{\alpha}e_{A}^{\alpha}\!=\!0$, and $N_{\alpha}k^{\alpha}\!=\!-1$ is given by
\begin{equation}
  N^{\alpha} = \frac{1}{2}\left(\frac{1}{A(r)B(r)},\frac{-1}{B(r)^{2}},0,0\right).
\end{equation}
Finally, we evaluate the ``transverse curvature'' \cite{2002gr.qc.....7101P} of $\Sigma$; this is defined as
\begin{align} \label{eq:TOVtransC}
  C_{ab} &\equiv -N_{\alpha}{e_{a}^{\alpha}}_{;\beta}e_{b}^{\beta} = \begin{pmatrix} \frac{A'(r)}{A(r)}+\frac{B'(r)}{B(r)} & 0 & 0 \\ 0 & \frac{-r}{2B(r)^{2}} & 0 \\ 0 & 0 & \frac{-r\sin^{2}\theta}{2B(r)^{2}} \end{pmatrix}. 
\end{align}

The stress-energy tensor on the hypersurface $\Sigma$ is given by
\begin{equation}
  T^{\alpha\beta}_{\Sigma}=(-k_{\mu}u^{\mu})^{-1}S^{\alpha\beta}\delta(\tau)\,,
\end{equation}
where $u^{\mu}$ is the 4-velocity of a congruence of timelike geodesics intersecting $\Sigma$, and $\tau$ is the proper time along the timelike geodesic congruence ($\tau=0$ is the intersection of the congruence with $\Sigma$), and
\begin{align}
  S^{\alpha\beta}=\mu k^{\alpha} k^{\beta} + j^{A}\left(k^{\alpha}e_{B}^{\beta}+e_{A}^{\alpha}k^{\beta}\right)+p\sigma^{AB}e_{A}^{\alpha}a_{B}^{\beta}, \label{eq:Tmunu}
\end{align}
with
\begin{align}
  \mu&=\frac{-1}{8\pi}\sigma^{AB}\left[C_{AB}\right]\,, \\ j^{A}&=\frac{1}{8\pi}\sigma^{AB}\left[C_{\lambda B}\right]\,, \\ \textrm{and }\, p&=\frac{-1}{8\pi}\left[C_{\lambda\lambda}\right]\,.\label{eq:endeq:Tmunu}
\end{align}
Square brackets denote the change of the enclosed quantity across $\Sigma$ from the past to the future, i.e.\ $\left[h(x)\right]=h(x)\big|_{\Sigma+}-h(x)\big|_{\Sigma-}$ \cite{2002gr.qc.....7101P}.
The function $\mu$ is a mass density on $\Sigma$, $p$ is an isotropic pressure on $\Sigma$, and $j^{A}$ represents a current on $\Sigma$ flowing in the $\theta$ and $\phi$ directions.

\begin{figure}[h]
\centering
\includegraphics[scale=0.50,clip]{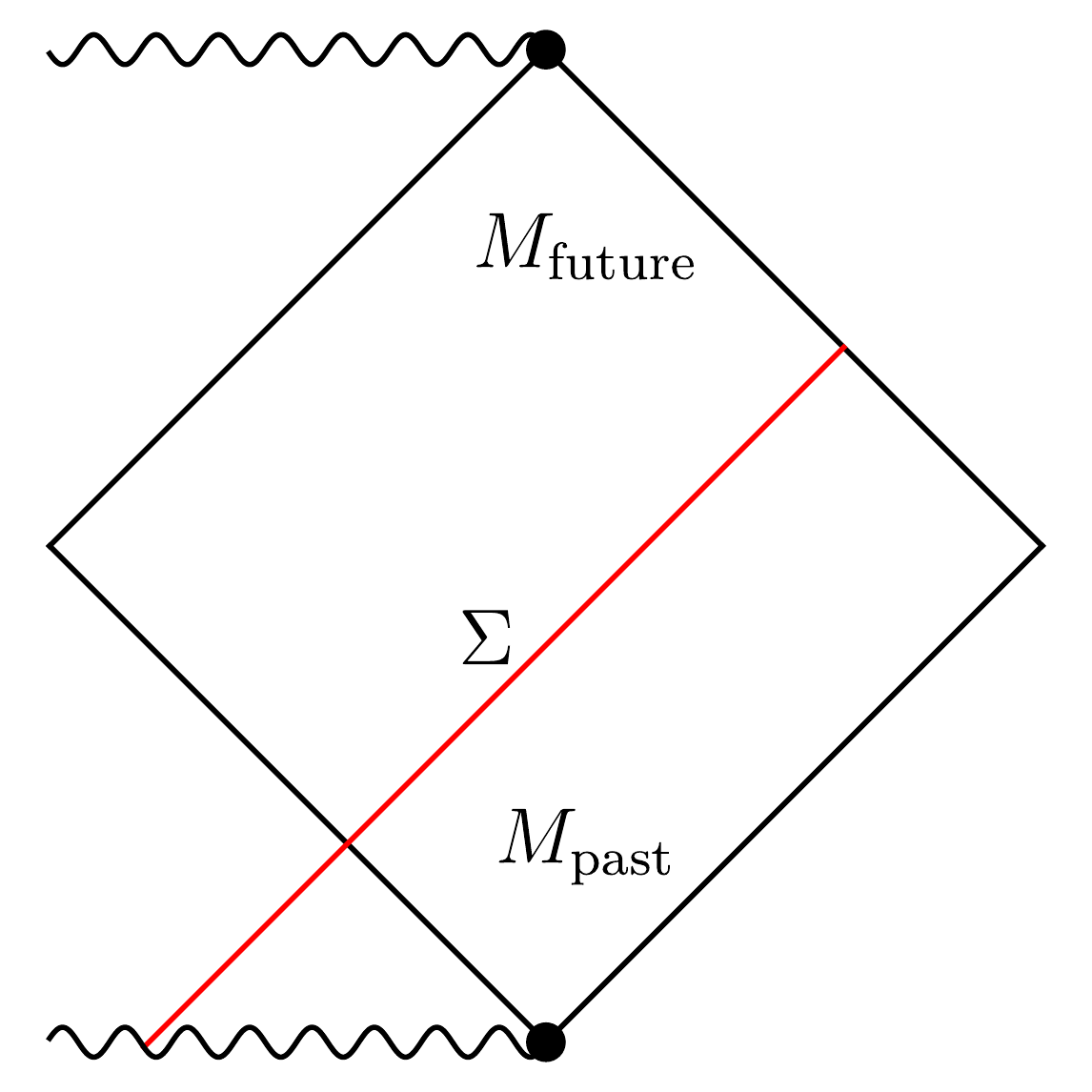}
\caption{\label{Fig_Schw} 
  The conformal diagram (or Penrose diagram) for the example of two Schwarzschild spacetimes joining along an outgoing null cone. This is a limiting case of the Vaidya spacetime \cite{Vaidya:1953zza}. The mass decreases across the null hypersurface, and the difference is radiated out to infinity by an expanding shell of null dust.
  }
\end{figure}

As an example, we first consider joining two Schwarzschild metrics with different masses along an outgoing null hypersurface, $\Sigma$. 
In the future [past] of $\Sigma$ we choose $A(r)^{2}=B(r)^{-2}={1-2M_{\textrm{future}}/r}$ $\big[A(r)^{2}=B(r)^{-2}={1-2M_{\textrm{past}}/r\big]}$.
Evaluating the above expressions we find that $\mu={(M_{\textrm{past}}-M_{\textrm{future}})/(4\pi r^{2})}$, ${p=0}$ and ${j^{A}=0}$.
Provided we choose $M_{\textrm{future}}<M_{\textrm{past}}$ then the stress-energy tensor of the matter on the null hypersurface describes a thin expanding shell of ``null dust'' containing a positive total energy $(M_{\textrm{past}}-M_{\textrm{future}})$.
The luminosity of this spacetime, as measured by an observer at future null infinity, is unbounded.
The spacetime we have constructed is a limiting case of the Vaidya spacetime \cite{Vaidya:1953zza} where the duration of the radiating region is taken to zero. 
The causal structure of this spacetime is shown in Fig.~\ref{Fig_Schw}; this spacetime is not a counter-example to the maximum luminosity conjecture in the main text, because the radiation comes out of a past horizon.

We now turn to the spacetime described in the main text, the metric is given in Eq.~\ref{eq:a}.
We define the following orthonormal tetrad, 
\begin{align}\label{eq:tetrad}
  W^{a} &= (f(r)^{-1/2},0,0,0)\,, \quad  X^{a} = (0,f(r)^{1/2},0,0)\,, \nonumber \\
  Y^{a} &= (0,0,r^{-1},0)\,,\quad  Z^{a} = (0,0,0,(r\sin\theta)^{-1})\,,
\end{align}
where $f(r)=1-2m(r)/r$. 
The stress-energy tensor in the bulk of this spacetime may be calculated from Einstein's equations, $T_{\mu\nu}=(1/8\pi){\cal G}_{\mu\nu}$, and decomposed onto the orthonormal tetrad in Eq.~\ref{eq:tetrad}. The stress-energy tensor is given by
\begin{equation}\label{eq:MYT}
  T^{\mu\nu} = \rho W^{\mu}W^{\nu} + p_{r} X^{\mu}X^{\nu} + p_{\theta} Y^{\mu}Y^{\nu}+ p_{\phi} Z^{\mu}Z^{\nu} \,, \\
\end{equation}
\begin{align}
  \textrm{with }\,\rho (r) &= -p_{r}(r) = \frac{3 a^2 M}{4 \pi  \left(a^2+r^2\right)^{5/2}} \,, \label{eq:NREF1}\\
  \textrm{and }\,p_{\theta} (r) &= p_{\phi} (r) = \frac{3 \left(3 a^2 r^2-2 a^4\right)M}{8 \pi  \left(a^2+r^2\right)^{7/2}} \,. \label{eq:NREF2}
\end{align}
%
t is straightforward to check that this stress-energy tensor satisfies the usual energy conditions \cite{Hawking:1973uf}. 
For example, the \emph{weak energy condition} states that that the inequality $T_{\mu\nu}X^{\mu}X^{\nu}\geq 0$ holds for any timelike vector $X^{\mu}$. For a stress energy tensor of the form in Eq.~\ref{eq:MYT} this condition will hold if $\rho\geq0$, $\rho+p_{r}\geq0$, $\rho+p_{\theta}\geq0$, and $\rho+p_{\phi}\geq0$, which can all be shown to hold from Eqs.~\ref{eq:NREF1} and \ref{eq:NREF2}. 
The \emph{null energy condition} states that that the inequality $T_{\mu\nu}X^{\mu}X^{\nu}\geq 0$ holds for any null vector $X^{\mu}$; the fact that the weak energy condition is satisfied implies (by continuity) that the null energy condition is also satisfied. 
Finally the \emph{dominant energy condition} states that for every timelike vector $X^{\mu}$ the inequality $T_{\mu\nu}X^{\mu}X^{\nu}\geq 0$ holds and $T_{\mu\nu}X^{\nu}$ is timelike. 
For a stress energy tensor of the form in Eq.~\ref{eq:MYT} this condition will hold if $\rho\geq 0$, $-\rho\leq p_{r}\leq\rho$, $-\rho\leq p_{\theta}\leq\rho$, and $-\rho\leq p_{\phi}\leq\rho$, which can all be shown to hold from Eqs.~\ref{eq:NREF1} and \ref{eq:NREF2}.

The transverse curvature of the null hypersurface, $\Sigma$, in this spacetime may be evaluated from equation \ref{eq:TOVtransC} and is given by 
\begin{equation} 
  C_{ab} =\begin{pmatrix} 0 & 0 & 0 \\ 0 & \left(\frac{M r^{3}}{4\left(a^2+r^2\right)^{3/2}}\!-\!\frac{r}{2}\right) & 0 \\ 0 & 0 & \left(\frac{M r^{3}}{4\left(a^2+r^2\right)^{3/2}}\!-\!\frac{r}{2}\right)\sin^{2}\theta \end{pmatrix}. 
\end{equation}
From equations \ref{eq:Tmunu} to \ref{eq:endeq:Tmunu}, the components of the stress energy tensor on, $\Sigma$, are given by
\begin{equation}
  \mu = \frac{\delta M r}{16 \pi  \left(r^2+1\right)^{3/2}} \,,\;
  p = 0 ,\; \textrm{and }\, j^{A} = 0\,. 
\end{equation}
This stress-energy tensor also satisfies all of the usual energy conditions and describes ``null dust''.
This is the result given in equation \ref{eq:delta} of the main text. 
The function $\mu(r)$ is plotted in Fig.~\ref{fig:Plummer}(b).

\end{document}